\documentstyle[emulateapj,psfig]{article}

\def\etal{{et~al.}}
\def\amin{\ifmmode^{\prime}\else$^{\prime}$\fi}
\def\asec{\ifmmode^{\prime\prime}\else$^{\prime\prime}$\fi}

\def\simgt{\lower.5ex\hbox{$\; \buildrel > \over \sim \;$}}
\def\simlt{\lower.5ex\hbox{$\; \buildrel < \over \sim \;$}}

\newcommand\xte{{\it RXTE}}

\newcommand\asca{{\it ASCA}}

\def\psr{\hbox{AX~J1845$-$0258}}
\def\src{\hbox{AX~J184453$-$025640}}
\def\dip{\hbox{AX~J184440$-$030501}}
\def\snr{\hbox{G29.6+0.1}}

\lefthead{Vasisht, Gotthelf, Torii, \&\ Gaensler}
\righthead{X-ray Observations of \psr}

\begin{document}

\title{Detection of a Compact X-ray Source in the Supernova Remnant G~29.6+0.1: A
       Variable Anomalous X-ray Pulsar?}

\author{G. Vasisht$^1$, E. V. Gotthelf$^2$, K. Torii$^3$, \&\ B. M. Gaensler$^4$}
\altaffiltext{1}{Jet Propulsion Laboratory,
California Institute of Technology, 4800 Oak Grove Drive, Pasadena,
CA, 91109, USA; gv@astro.caltech.edu} \altaffiltext{2}{Columbia
Astrophysics Laboratory, Columbia University, 550 West 120$^{th}$
Street, New York, NY 10027, USA; evg@astro.columbia.edu}
\altaffiltext{3}{NASDA TKSC SURP, 2-1-1 Sengen, Tsukuba, Ibaraki 305-8505
Japan; torii.kenichi@nasda.go.jp}                                     
\altaffiltext{4}{Hubble Fellow, Center for Space Research,
Massachusetts Institute of Technology, Cambridge, MA, 02139;
bmg@space.mit.edu}

\begin{abstract} 

We present follow-up observations of the serendipitously discovered
7-s x-ray pulsar \psr, which displays characteristics similar to those
observed in the anomalous x-ray pulsars (AXPs). We find a dramatic
reduction in its 3-10 keV flux in both new ASCA and RXTE
datasets. Within the pulsar's position-error locus, we detect a faint
point source, \src, surrounded by an arc of diffuse x-ray
emission. This arc is coincident with the South-East quadrant of the
radio shell of the newly discovered supernova remnant \snr, reported
in our companion paper (Gaensler et al. 1999). Lack of sufficient
flux from the source prevents us from confirming the 7-s pulsed
emission observed in the bright state; hence, at present we cannot
definitively resolve whether \psr\ and \src\ are one and the same. If
they are the same, then the peak-to-peak luminosity changes recorded
for \psr\ may be larger than seen in other AXPs; closer monitoring of
this pulsar might lead to a resolution on the mechanism that drives AXPs.

\end{abstract}


\section{Introduction}

AX~J1845$-$0258 is a 7-s x-ray pulsar discovered during an automated
search of the ASCA archival data (Gotthelf \& Vasisht 1998, herein
GV98; Torii et al. 1998). The pulsar lies $22^{\prime}$ away from the
supernova remnant Kes~75, the main target of that ASCA pointing; this
large angular separation makes an association between Kes~75 and the
pulsar highly improbable.  Arguing on the basis of its spectral and
timing properties, we proposed \psr\ to be the latest addition to the
class of anomalous x-ray pulsars (AXPs; GV 98; Torii et al. 1998 - for
AXP phenomenology see Mereghetti \& Stella 1995 and van Paradijs et
al. 1995, for the AXP-magnetar interpretation see Thompson \& Duncan
1996). The collective evidence included the long rotation period,
large sinusoidal pulse modulation, steady x-ray flux on timescales of
a day or less, and a soft power-law x-ray spectrum.  A rough distance
estimate derived from the large line-of-sight x-ray absorption placed
the pulsar at a distance of 5 to 15 kpc, with an inferred x-ray
luminosity of $\sim 2.5 \times 10^{35}d_{15}^2$ erg s$^{-1}$ (the
distance being $15d_{15}$ kpc), not atypical for AXPs.

Since the small AXP population shows a propensity towards association
with supernova remnants we undertook searches for a host supernova
remnant at radio and x-ray wavelengths.  In this letter, we report
follow-up ASCA \& RXTE x-ray observations targeted at the pulsar. In
our companion paper, we reported on a VLA detection of a young radio
shell coincident with the pulsar's error circle (Gaensler, Gotthelf \&
Vasisht 1999; hereafter GGV99). The primary goal of our x-ray
observations was to identify the pulsar and confirm or repudiate the
AXP hypothesis by measuring the spin-down rate of the pulsar.  As in
the radio, we succeeded in finding evidence of a young x-ray SNR
within the pulsar's error circle, however, pulsed emission was not
observed in these followup observations. Instead, we find a faint ASCA
point source at the center of the newly discovered radio remnant \snr\
(GGV99).  We argue that this faint source is indeed the pulsar \psr\
albeit in a low state, and that its location in the center of a
supernova remnant additionally favors the anomalous x-ray pulsar
interpretation. The angular size of \snr\ and limits on its distance
suggest that the original detonation is no more than 8-kyr old (see
GGV99). This implies that the slow rotator at the remnant's core could
well be a spun-down magnetar.

\section{Observations and Analysis}

A new x-ray observation of the field containing the pulsar \psr\ was
obtained with the Advance Satellite for Cosmology and Astrophysics
(ASCA; Tanaka et al. 1994) on March 28-29, 1999 (UT), with
with both pairs of on-board instruments, the two solid-state
spectrometers (SIS) and the two gas imaging spectrometers (GIS). 
The SIS data were
in 1-CCD mode with the pulsar centered as close to the mean
SIS telescope optical axis as was practical, to minimize vignetting
and off-axis aberrations.  The spatial resolution for the SIS
is limited by the optics to $\sim 1^{\prime}$,
while the GIS spatial resolution of $2-3^{\prime}$ is due to an
additional energy dependent instrumental blur.  The GIS data were
collected in the highest time resolution mode ($0.5 \ \rm{ms}$ or $64
\ \mu\rm{s}$, depending on the telemetry rate), with reduced spectral
binning ($\sim 47$ eV per PHA channel).  All data were edited to
exclude times of high background contamination.
The resulting effective observation time was
$49$~ks ($64$~ks) for each GIS (SIS) sensor.
 
Figure 1 displays the broad-band ($1-10$ keV) GIS image of the pulsar
field produced by co-adding exposure corrected maps from both
instruments smoothed using a 3$^2$ pixel box-car filter.  Near the
center of the image lies a faint ASCA source (marked by a cross),
confined to the original 3 arcminute error circle for \psr\ (GV98).
Inspection of the higher spatial resolution SIS image (figure 2a)
resolved this emission into a
$1^{\prime}$ SIS point-source surrounded by a diffuse arc of x-rays,
just south-east of the point-like emission. The arc coincides with the
$4^{\prime}$ diameter radio shell of the recently discovered supernova
remnant \snr\ (see GGV99), and overlaps the sector where the
radio emission is the strongest.  The location of the SIS
source at the center of \snr\ makes a physical association between the
two very likely. The lack of a complete x-ray shell with
correspondence to the radio remnant is not unexpected, considering the
high foreground absorption associated with this region.
The identification of the central source in the SIS allows a
coordinate determination of $18^h 44^m 53^s$, $-02^{\circ} 56^{\prime}
40^{\prime\prime}$ (J2000) with an uncertainty of $12^{\prime\prime}$
radius. This reduced ASCA error circle is derived using the method
developed to compensate for the temperature dependent star tracker
drift in the aspect solution (Gotthelf \etal\ 2000).  Herein we
refer to this source as \src, and will consider in detail (\S 3)
whether this is indeed \psr, albeit at a lower flux level.
 
The source count rate at the putative pulsar location is evidently
reduced in the 1999 observation. 
The background-subtracted source count-rate in the optimal
$3-10$ keV energy band is $4.0 \pm 0.7 \times 10^{-3}$ s$^{-1}$
(combined SIS) after correcting for aperture losses, resulting in a
$6$ $\sigma$ detection. For an invariant pulsar (based on the 1993
dataset), the expected count
rate would be $3.9 \times 10^{-2}$ s$^{-1}$. Thus we can place a limit
on the variability of a factor of 9.7 in flux between the 1993 and
1999 observation epochs, assuming the spectral shape has remained
unchanged (see discussion). This low flux level (in 1999) is consistent with the
marginal non-detection of the pulsar in a short (10-ks) 1997 ASCA
observation of the Galactic ridge reported by Torii \etal\ (1998)
suggesting that the pulsar was in a similar low state then.
 
In addition to the compact source and the new supernova remnant \snr,
and the well studied supernova remnant Kes 75 at the eastern edge of
the GIS field-of-view, two additional objects are evident in the GIS
image of Figure 1.  Towards the southwest of \snr, lies a moderately
bright unresolved GIS point-source, which we name \dip\ based on its
$2^{\prime}$ GIS position. The source spectrum containing 550 counts,
is fit well by an absorbed power law of photon index 1.7 with an N$_H
= 6 \times 10^{22}$ cm$^{2}$.  The unabsorbed flux is $3 \times
10^{-12}$ ergs cm$^{-2}$ s$^{-1}$ ($2-10$ keV).  Although this source
is located off our SIS and VLA maps, examination of databases of this
region shows that \dip\ lies at the edge of a complex radio/IR region
cataloged variously as GRS 029.39+00.10/PMN J1844-0306/F3R 1015. No
other cataloged object in any wavelength is recorded for these
coordinates. We also find hard diffuse emission that is offset $\sim
10'$ due northeast of the remnant. The morphology is of an extended
bow-tie-like structure, that could well be a partial shell SNR.  We
found no cataloged counterparts for this object. Table 1 summarizes
the positions of these sources.

We searched the GIS data for evidence of pulsed emission from \src\
around the 7-s period. A total of $1,418$ photons from the two GISs
were extracted from a $8^{\prime}$ diameter aperture and merged of
which $\sim 300$ counts are expected from the compact source. The
photon arrival times for each event were corrected to the solar system
barycenter. The data were then folded in period space around the 1993
value with a range to accommodate spin-up or down values ($| \dot P |
\le 1 \times 10^{-10}$ s s$^{-1}$) in $0.1 \times P^2 / 2 T$ steps, in
order to search for a coherent modulation.  No significant period was
found. We place a limit of 0.7 on the fractional 
modulation (for a 5$\sigma$ pulse detection threshold), higher 
than the modulation of 0.3 found in the discovery observations of 1993. 
We also searched for a signal from \dip\ in the range $0.02-500$ s, but found
no significant periodicity.

We also observed the region containing the pulsar using the \xte\
observatory on 18 Apr 1999.  Data was acquired with the Proportional
Counter Array (PCA) in ``Good Xenon'' data mode at 0.9 $\mu$s time
resolution. The PCA instrument covers an energy range of $1-40$ kev
with an effective area of $6,400$ cm$^2$ over its $\sim 1^{\circ}
\times 1^{\circ}$ field-of-view (FWHM). After processing and
barycentering the Good Xenon data, we
obtained a total of 38 ks of screened exposure time. Photons were
further restricted to the energy range $\simlt 10$ keV from layer 1
only.  Since the \xte\ data was nearly contemporaneous with the \asca\
observation, we expect the pulsar to be in the low state. In any case,
we searched for the pulsar in a manner similar to that for the \asca\
data using data below 6 keV to cut off excess background. 
No significant periodicities (above 5 $\sigma$) were found. Given the
count rate expected in the low state, no useful modulation limits may
be set with this dataset.  

\section{Discussion}

The failure to detect a pulsed source clearly corresponding to \psr\
was somewhat surprising in the light of its interpretation as an AXP,
but is not completely confounding.  At the same time, the discovery of
the young supernova remnant \snr\ at that position does help bolster
the AXP interpretation. The detection of the fainter point source,
\src, at the core of the remnant and its co-location with the error
locus of \psr\ strongly suggests that the two compact sources are one
and the same.  Our conclusion is that the pulsar must have
undergone a factor-of-ten variability in measured flux interim to the
ASCA epochs spanning six years.  This behavior is somewhat unusual.
There is evidence to show that the two well studied AXPs,
1E~1048.1-593 and 1E~2259+586, display about a factor-of-four flux
variations on year long timescales (Baykal \& Swank 1996; Oosterbroek
et al. 1998).  The compact source in the core of RCW 103 also displays
order-of-magnitude flux variations (Gotthelf, Petre \& Vasisht 1999)
in the $3 - 10$ keV band. The latter, although showing many of the
same properties as AXPs, is not classified as an one because of
absence of pulsed emission (but see Garmire et al. 2000). Other
objects show more steady behavior - the pulsar in Kes~73 shows fluxes
that are steady to within a factor-of-two over a decade of monitoring.

In GV98 we argued that the x-ray spectrum of \psr\ is in all
likelihood, highly absorbed thermal emission which is observed as a
steep power-law (photon index $\simeq 5$ with $N_H \sim 10^{23}$
cm$^{-2}$) above 2 keV, with few photons below that energy.  The
emission may be modeled as surface blackbody radiance with temperature
$kT \sim 0.6 \pm 0.1$ keV.  When coupled with the flux, this implies a
hotspot-like emitting region of fraction $0.1 d_{15}^2$ of the total
surface area of a neutron star of standard radius; this estimate is
rough - it assumes isotropic emission and ignores any relativistic
corrections. Given the BB model, it is important to clarify that
the large foreground absorption can lead to observed flux variations 
of greater magnitude than actual intrinsic variations. For instance
surface cooling on the NS (at 0.6 keV) leading to an intrinsic flux change
of a factor-of-five, can lead to a factor-of-ten flux change in the
observed Wien tail, given $10^{23}$ cm$^{-2}$ of foreground absorption.

This is only the third convincing association of an AXP
with a supernova remnant (GGV 99). The other two cases are the
associations of 1E~2259+586 with CTB 109 (Gregory \& Fahlman 1980) and
1E~1851$-$045 with Kes 73 (Vasisht \& Gotthelf 1997). The handful of
other AXPs are not known to be associated with bright supernova
remnants, which suggests that these are somewhat older objects
(although, these field need to be imaged with greater sensitivity). If
ultramagnetized, then AXPs must be the youngest observable magnetars,
spanning about three decades in
age grouping. There is strong observational evidence that 1E
1851$-$045 in Kes 73 is no more than $4\times 10^3$ yr-old from its
timing parameters, while dating of Kes 73 suggests that the PSR/SNR pair is
perhaps as young as $2\times 10^3$ yr (Vasisht \& Gotthelf 1999).
Anomalous pulsars that have no obvious SNR counterparts are dated to
be $\sim 10^5$ years in age. Implicit in the previous statement, is
the assumption that the spindown-ages of AXPs reflect their
true ages. Field decay and wind induced torques may result in
significant departures the age and its estimator,
especially for the older objects.  Beyond the $10^5$ yr timescale,
magnetic activity, which is believed responsible for powering the
persistent emission in these objects (Heyl \& Kulkarni 1998) 
declines rapidly, leading to their disappearance from the observable
x-ray sky - resulting is an estimated $10^8$ defunct Galactic
magnetars.  An example of elderly magnetar may
be the nearby, slow x-ray pulsar RX~J0720.4$-$3125 (Haberl et
al. 1997; Kulkarni \& van Kerkwijk 1998). The soft
$\gamma$-ray repeaters may well be a phase in the life of AXPs (for
instance, Gaensler 1999), lasting for about $10^{4-5}$ years, and
triggered by a yet ill-understood mechanism. 

If \psr\ is indeed a magnetar, then the magnetar mechanism must
address the cause of the x-ray flux variability. In alternative
accretion scenarios (which do not require invoking ultramagnetized
stars), changes in $L_X$ track the mass accretion rate onto the star.
We have observed flux variability of at least a factor-of-five,
possibly accompanied by a change in the emission temperature (we are
unable to confirm that with the current data), which exists on
timescales of a few years or less. Although no short term variability
or stochastic flickering, seen in accreting objects, is observed in
the lightcurves of \psr, a plunge in the mass accretion rate $\dot
M_X$ from an ejecta-fallback disk can easily account for the drop in
$L_X$ (see Chatterjee, Hernquist \& Narayan 1999, Alpar 1999, Marsden
et al. 1999). We do not know what mechanism sets a variability
timescale of $\sim$ years in the otherwise steady $\dot M_X$ decline
of a fallback disk.  In the magnetar picture, variations in the
surface flux will be driven by and will track the magnetic field
dissipation inside the star. The shape of the stellar spectrum,
inferred temperatures and emitting area ($\simeq 0.1 A_s$) suggest
thermal activity related to a heated spot on the stellar surface; this
is natural as a strong field suppresses thermal conductivity
perpendicular to the field vector (Hernquist 1985). The thermal
conduction time in a neutron star (of surface $kT \sim 0.5$ keV), from
core to surface, is about a year (Van Riper, Horn \& Miller 1991),
which smoothes out any surface temperature variations
on timescales shorter than about a year. Detectable changes in the
surface flux will be driven only by longterm internal dissipation
cycles. For \psr\, the total energy released in a longterm event is
about $\delta E \sim 10^2\tau_{var} L_X \sim 2 \times 10^{44}$ erg,
the factor $10^2$ allows for most of the cooling to be in the form
of neutrino emission (Thompson \& Duncan 1996). A magnetar has a
magnetic free energy budget to go through $10^3$ dissipation events of
the given magnitude during the $\sim 10^4$ yr lifetime of \psr.
Another mechanism for variability in a magnetar could be radiative
precession (with a few year timescale) of the spin axis of the star
around the dipole axis if these are significantly misaligned; this
results from hydrodynamic deformation induced in the star by the
strong B-field (Melatos 1999).

Finally, if variability is a common aspect of the younger anomalous
x-ray pulsars, then that raises a couple of interesting questions.  If
AXPs are magnetars (or drawn from a population {\it P}), and AXPs show flux
variability, then strictly, is it likely that we have significantly
under-counted them and therefore underestimated the birth rate of
magnetars (or {\it P}s), estimated to be $\sim 10^{-3}$ yr$^{-1}$?  Is it
possible that there are such neutron stars within some of the young
Galactic remnants that apparently have no compact objects or plerionic
cores associated with them, as the recent discovery of the low $L_X$
compact source in Cas A (Pavlov et al. 1999, Chakrabarty et al. 2000)
might suggest? The answers to these questions are unclear at
present. First, we have no knowledge whether flux variations can be
greater than an order of magnitude (with the current population of
well monitored AXPs, that appears not to be the case). Second, we do
not know the duty cycle of these variations. In the case of
1E~2259+586 and 1048.1$-$593 it seems to be of order a year.  Thirdly,
AXPs are intrinsically bright sources ($L_X \simeq 10^{35}$ erg
s$^{-1}$), with the consequence that mild variability is unlikely to
lead to significant under-counting. And finally, the large distances
to the known population of objects and the fact that no AXP has been
found nearby ($\sim 1$ kpc), strongly suggests that these objects are
indeed rare. At the moment there is little evidence to link low $L_X$
objects such as the compact sources in Cas A, Puppis A and PKS
1209$-$51 to the brighter AXP population.  In any case, further
monitoring of the levels of variability and timescales involved will
lead to a better comprehension of any physical ties among these
categories of sources.  The physical mechanism driving the x-ray
luminosity in this object (and by extension, other AXPs) could well be
pinned down through monitoring the spin evolution of this pulsar
(\psr\ may undergo additional large peak-to-peak episodes in $L_X$, or
make a recovery to the flux levels observed in the 1993 ASCA
observation) through episodes of variability with missions such as
Chandra and XMM, and observing how the stellar period tracks any
changes in $L_X$.

\begin{acknowledgements}
{\noindent \bf Acknowledgments} --- E.V.G. \& G.V. acknowledge the
support of NASA LTSA grant NAG5--7935.  B.M.G.  acknowledges the
support of NASA through Hubble Fellowship grant HF-01107.01-98A
awarded by the Space Telescope Science Institute, which is operated by
the Association of Universities for Research in Astronomy, Inc., for
NASA under contract NAS 5--26555.
\end{acknowledgements}


\footnotesize
\begin{deluxetable}{lccc}
\tablewidth{300pt}
\tablecaption{Coordinates of Field X-ray Sources
\vfill \label{tbl-1}}
\tablehead{
\colhead{Object} & \colhead{R.A.} & \colhead{DEC}& \colhead{Uncertainty} \nl
      \colhead{} & \colhead{(J2000)}    & \colhead{(J2000)}    & \colhead{}    
}
\startdata
\src\                 & $18^h 44^m 53^s$ & $-02^{\circ} 56^{\prime} 40^{\prime\prime}$ & $20^{\prime\prime}$\nl
Southeast source  & $18^h 44^m 40^s$ & $-03^{\circ} 05^{\prime} 01^{\prime\prime}$& $1^{\prime}$\nl
Center of ``Tie'' & $18^h 45^m 27^s$ & $-02^{\circ} 50^{\prime} 02^{\prime\prime}$& $2^{\prime}$\nl
Northern  ``Tie''  & $18^h 45^m 15^s$ & $-02^{\circ} 47^{\prime} 50^{\prime\prime}$& $2^{\prime}$\nl
Southern  ``Tie''  & $18^h 45^m 38^s$ & $-02^{\circ} 51^{\prime} 52^{\prime\prime}$& $2^{\prime}$\nl
\enddata
\end{deluxetable}


\begin{figure}[here]
\centerline{\psfig{file=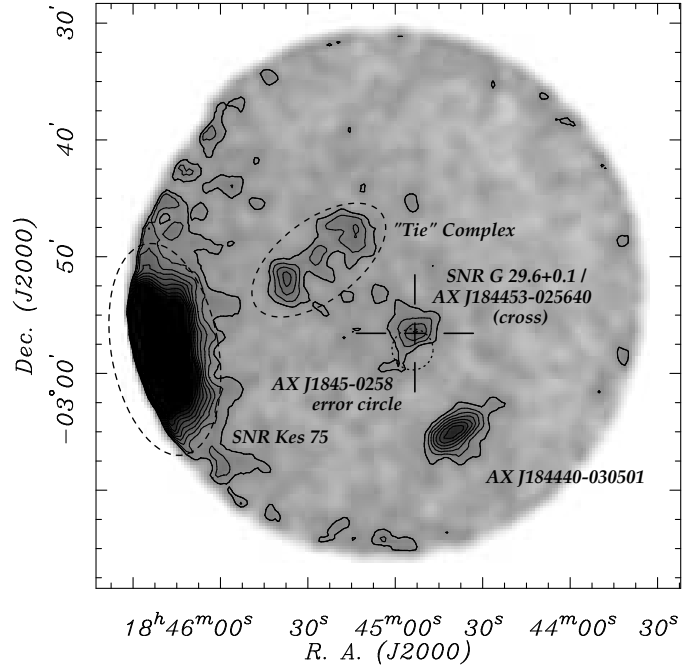,height=3.5in,angle=270,clip=}}
\caption[]{\footnotesize The full field-of-view of the 1999 ASCA GIS
observation of \psr. The processed image shows the new x-ray source
\src\ marked by the cross which lies within the error circle for \psr\
(dotted circle), and the two previously uncataloged x-ray sources to
the northeast and southwest. The bright emission towards the eastern
edge is the SNR Kes 75.}
\end{figure}

\begin{figure}[here]
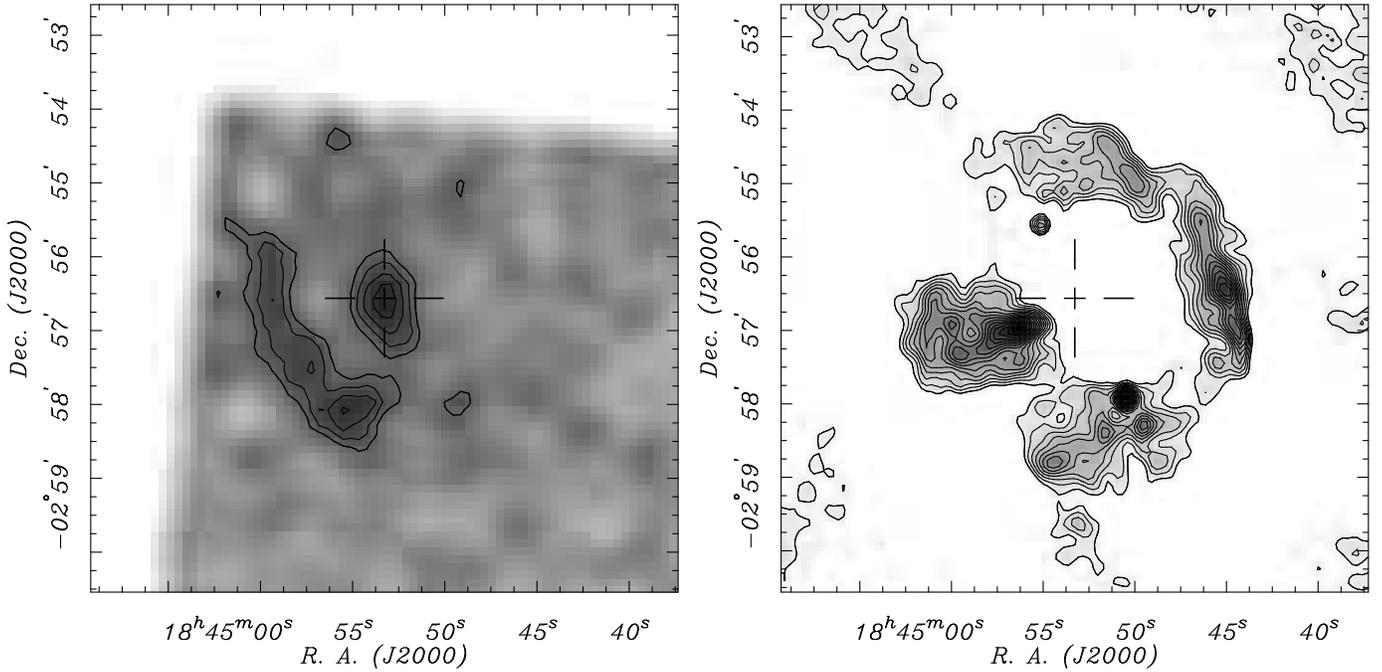

\centerline{
\psfig{file=ax1845_paper2_fig2a.ps,height=3.5in,angle=270,clip=} \hfil
\psfig{file=ax1845_paper2_fig2b.ps,height=3.5in,angle=270,clip=}
}

\caption[]{\footnotesize The new SNR, \snr, containing the compact
x-ray source, \src, within the error box of the x-ray pulsar \psr.
({\bf LEFT}) The ASCA SIS x-ray image centered on the \src, marked by
the cross. An arc of emission surrounds the point source and overlaps
with the radio shell displayed in the adjacent panel. {\bf (RIGHT)} A
5 GHz VLA map of the same region, illustrating the clumpy shell
\snr. Again, the pulsar's location is marked by the cross.  }
\end{figure}

\end{document}